# An approximate simple equation of state for model fluids

by Richard BONNEVILLE

Centre National d'Etudes Spatiales (CNES), 2 place Maurice Quentin, 75001 Paris, France

phone: +33 1 44 76 76 38, mailto:richard.bonneville@cnes.fr

**Abstract**

A new analytical approach to derive an approximate equation of state and the virial coefficients for simple fluids is presented. Starting from the usual expression of the partition function, we first perform a Fourier transformation, and then we implement an original approximation scheme of the correlation terms in the reciprocal space. A procedure involving combinatorial analysis and partial re-summations leads to simple expressions for the virial coefficients, which are effectively calculated for hard sphere fluids. A comparison is made with Carnahan & Starling's equation of state and with the available numerical results for the virial coefficients. The virial coefficients calculated within our approximation scheme are in rather good agreement with the known computed values up to order 6, i.e. at low density; if we extrapolate those results for higher orders we obtain a simple equation of state in good agreement with the experimental data for the stable fluid phase.

## Introduction

So far an enormous quantity of theoretical as well as numerical and experimental works has been dedicated to proposing equations of state for fluids [1]. The special case of hard sphere fluids has been all the more investigated as the random packing of a large number of identical hard spheres is a fruitful model for simple fluids [2]. Quite generally the equation of state can be expressed as an expansion in powers of the density, the "virial expansion":

$$\frac{PV}{Nk_BT} = 1 + B_2\rho + B_3\rho^2 + B_4\rho^3 + ... \tag{1}$$

$P$ is the pressure, $k_B$ is the Boltzmann constant, $T$ is the absolute temperature, $N$ is the number of particles and $V$ is the volume; $N$ and $V$ are arbitrary large, whereas the density $\rho = N/V$ is a finite parameter. The $B_p$s are the "virial coefficients"; they can be derived from theoretical equations of state or computed by numerical simulations.

In a previous paper [3], we had presented a new analytical approach to the computation of the virial coefficients for simple fluids based on working in the reciprocal space, with a focus on the hard sphere case. We here under present an approximation scheme extending that approach so as to explicitly derive an equation of state and the virial coefficients. Starting from the usual expression of the partition function of a set of identical molecules we will first perform a Fourier transformation. We will then implement an original approximation of the correlation terms in the reciprocal space. A procedure involving combinatorial analysis and partial re-summations will lead to simple expressions for the virial coefficients, which will be effectively calculated for hard sphere fluids. A comparison will be made with Carnahan & Starling's equation of state [4] and with the available numerical results for the virial coefficients e.g. by Clisby & Mc Coy [5].

## The partition function in the reciprocal space

In this section we summarize the process detailed in ref.[3]. We first assume that the intermolecular potential only includes pair interactions and we also assume that the interaction potential $U(\mathbf{R}_p - \mathbf{R}_q)$ only depends upon the relative position of the molecules. We let the internal degrees of freedom aside. The starting point is the expression of the semi classical partition function of an assembly of identical molecules of mass $M$:

$$\mathbb{Z} \cong \frac{(2\pi M k_B T)^{3N/2}}{\hbar^{3N}} \frac{V^N}{N!} Z \tag{2}$$

where

$$Z = V^{-N} \int_{(V)} d^3\mathbf{R}_1 \int_{(V)} d^3\mathbf{R}_2 \ldots \int_{(V)} d^3\mathbf{R}_N \exp\left(\frac{-1}{k_B T} \sum_{p>q} U(\mathbf{R}_p - \mathbf{R}_q)\right) \tag{3}$$

The pressure is derived as

$$P = k_B T \frac{\partial \text{Log}(\mathbb{Z})}{\partial V} \tag{4}$$

We first perform a Fourier transformation by writing:

$$\Phi(\mathbf{k}_{pq}) = \int d^3\mathbf{R} \left[\exp(-U(\mathbf{R}_p - \mathbf{R}_q)/k_B T) - 1\right] \exp(-i\mathbf{k}_{pq} \cdot \mathbf{R}) \tag{5}$$

and in a reciprocal way

$$\exp(-U(\mathbf{R}_p - \mathbf{R}_q)/k_B T) = \int \frac{d^3\mathbf{k}_{pq}}{(2\pi)^3} \left[(2\pi)^3 \delta(\mathbf{k}_{pq}) + \Phi(\mathbf{k}_{pq})\right] \exp(i\mathbf{k}_{pq} \cdot (\mathbf{R}_p - \mathbf{R}_q)) \tag{6}$$

No assumption is made at this stage on the precise shape of the intermolecular pair potential $U(\mathbf{R}_p - \mathbf{R}_q)$. If we restrict ourselves to the case of hard spheres of diameter $d$, i.e. $U(\mathbf{R}_p - \mathbf{R}_q) = 0$ for $|\mathbf{R}_p - \mathbf{R}_q| > d$ and $U(\mathbf{R}_p - \mathbf{R}_q) = +\infty$ for $|\mathbf{R}_p - \mathbf{R}_q| < d$, then

$$\Phi(\mathbf{k}) = -\frac{4\pi}{k^3}\left(\sin(kd) - kd\cos(kd)\right) \tag{7}$$

From equ.(3) and equ.(5) $Z$ can be written as:

$$Z = \prod_{p>q} \int \frac{d^3\mathbf{k}_{pq}}{(2\pi)^3} \left[(2\pi)^3 \delta(\mathbf{k}_{pq}) + \Phi(\mathbf{k}_{pq})\right] \Delta \tag{8}$$

We have put

$$\Delta = \prod_m \int_{(V)} \frac{d^3\mathbf{R}_m}{V} \exp\left(i\mathbf{R}_m \cdot \sum_n \mathbf{k}_{mn}\right) \text{ with } \mathbf{k}_{mn} = -\mathbf{k}_{nm} \tag{9}$$

Now

$$\lim_{V \to \infty} \Delta = \prod_m \frac{(2\pi)^3}{V} \delta\left(\sum_{n \neq m} \mathbf{k}_{mn}\right) \qquad (10)$$

The dimensionless term $\Delta$ accounts for the intermolecular correlations; its value is 1 if all the arguments of the $\delta$s are null, otherwise it is 0.

The $\{\mathbf{k}_{mn}\}$s can be represented by an anti-symmetric $N \times N$ matrix whose sums of the lines and of the columns are zero.

In the mean field approximation where all correlations between the $\mathbf{k}_{mn}$ are discarded, that matrix is identically null; $\Delta$ is then approximately written as

$$\Delta \cong \prod_{m>n} \frac{(2\pi)^3}{V} \delta(\mathbf{k}_{mn}) \qquad (11)$$

**Beyond the mean field approximation**

In order to go beyond the mean field case, the idea is to propose a pertinent, less crude approximation for $\Delta$. For that purpose we hereunder use the following approximate form for the $\{\mathbf{k}_{mn}\}$ matrix:

$$\begin{pmatrix} 0 & +\mathbf{q} & 0 & 0 & 0 & \ldots & 0 & 0 & 0 & 0 & -\mathbf{q} \\ -\mathbf{q} & 0 & +\mathbf{q} & 0 & 0 & \ldots & 0 & 0 & 0 & 0 & 0 \\ 0 & -\mathbf{q} & 0 & +\mathbf{q} & 0 & \ldots & 0 & 0 & 0 & 0 & 0 \\ 0 & 0 & -\mathbf{q} & 0 & +\mathbf{q} & \ldots & 0 & 0 & 0 & 0 & 0 \\ \ldots & \ldots & \ldots & \ldots & \ldots & \ldots & \ldots & \ldots & \ldots & \ldots & \ldots \\ \ldots & \ldots & \ldots & \ldots & \ldots & \ldots & \ldots & \ldots & \ldots & \ldots & \ldots \\ 0 & 0 & 0 & 0 & 0 & \ldots & -\mathbf{q} & 0 & +\mathbf{q} & 0 & 0 \\ 0 & 0 & 0 & 0 & 0 & \ldots & 0 & -\mathbf{q} & 0 & +\mathbf{q} & 0 \\ 0 & 0 & 0 & 0 & 0 & \ldots & 0 & 0 & -\mathbf{q} & 0 & +\mathbf{q} \\ +\mathbf{q} & 0 & 0 & 0 & 0 & \ldots & 0 & 0 & 0 & -\mathbf{q} & 0 \end{pmatrix} \qquad (12)$$

All the elements of the diagonal line on the upper (resp. the lower) side of the main diagonal have the same value $+\mathbf{q}$ (resp. $-\mathbf{q}$), all the other elements are zero, except $\mathbf{k}_{1N} = -\mathbf{q}$ and $\mathbf{k}_{N1} = +\mathbf{q}$ in order to respect the boundary conditions; an equivalent way would be to call for cyclic conditions with particle N+1 = particle 1.

Z as given by equ.(7) can then be expanded as a series of the density; a systematic procedure involving combinatorial analysis and partial re-summations analogous to [5] finally allows expressing Z as

$$Z \cong \left(1 + \frac{\Phi(0)}{V}\right)^{N(N-1)/2} \exp N\left(\frac{N^2 I_3}{3!} + \frac{N^3 I_4}{4!} + \frac{N^4 I_5}{5!} + \ldots\right) \qquad (13)$$

The first bracket on the right hand side is the mean-field partition function $Z_{MF}$. As $\Phi(0)$ is of the order of a few molecular volumes and since $\lim_{N \to \infty}\left(1 + \frac{x}{N}\right)^N = \exp(x)$, we get

$$Z_{MF} \cong \exp\left(N \frac{\rho \Phi(0)}{2}\right) \qquad (14)$$

The $I_p$ are integral expressions given by

$$I_p = \int \frac{V d^3\mathbf{q}}{(2\pi)^3}\left(\frac{\Phi(q)}{V}\right)^p \times \int \frac{V d^3\mathbf{q}_1}{(2\pi)^3}\left(\frac{\Phi(q_1)}{V}\right) \times \int \frac{V d^3\mathbf{q}_2}{(2\pi)^3}\left(\frac{\Phi(q_2)}{V}\right) \times \ldots \times \int \frac{V d^3\mathbf{q}_{p-1}}{(2\pi)^3}\left(\frac{\Phi(q_{p-1})}{V}\right) \qquad (15)$$

The resulting equation of state is

$$\frac{PV}{Nk_B T} \cong 1 - \frac{\rho \Phi(0)}{2} - \frac{N^2 (2I_3)}{3!} - \frac{N^3 (3I_4)}{4!} - \frac{N^4 (4I_5)}{5!} + \ldots \qquad (16)$$

It is easily shown that for hard spheres

$$\int \frac{V d^3\mathbf{q}}{(2\pi)^3}\left(\frac{\Phi(q)}{V}\right) = -1 \qquad (17)$$

We assume that this relation is also valid for real fluids so that

$$I_p = (-1)^{p-1} \int \frac{V d^3\mathbf{q}}{(2\pi)^3}\left(\frac{\Phi(q)}{V}\right)^p \qquad (18)$$

The virial coefficients as defined by equ.(1) are thus given by

$$B_p = \frac{(-1)^p (p-1)}{p!} \int \frac{d^3\mathbf{q}}{(2\pi)^3}(\Phi(q))^p \qquad (19)$$

In the hard sphere case, introducing the molecular volume $v = \pi d^3/6$ and the packing density parameter $\xi = Nv/V$ we have

$$\Phi(0) = \lim_{k \to 0} \Phi(\mathbf{k}) = -\frac{4\pi}{3}d^3 = -8v. \tag{20}$$

and

$$I_3 = -30(v/V)^2 \tag{21}$$

The following terms can be computed numerically. In function of the packing density parameter $\xi$ the virial coefficients for hard spheres are given by

$$B_p = \frac{-(p-1)}{p!}I_p \bigg/ \left(\frac{v}{V}\right)^{p-1} \tag{22}$$

We explicitly find for the first terms

$$\frac{PV}{Nk_BT} \cong 1 + 4\xi + 10\xi^2 + 20.72\xi^3 + 32.50\xi^4 + 42\xi^5 + \ldots \tag{23}$$

**Discussion**

So it seems that the agreement with the known computed values is rather good up to order 6, but in fact beyond $p = 6$ our $B_p$ s show a maximum for $p = 7$ and then they decrease and finally go down to 0 (we get $B_{20} \cong 3 \times 10^{-2}$). That behaviour is actually not surprising: the more the density increases the more the higher order term of the virial expansion become important and our approximation becomes less and less valid. Now, we know that at high density the analytical behaviour of the virial expansion changes; the Carnahan & Starling equation which exhibits a pole for $\xi = 1$ is no longer correct and the equation of state should adopt the following asymptotic behaviour, valid in the metastable phase [6]

$$\lim_{\xi \to \xi_0} \frac{PV}{Nk_BT} = \frac{A}{1 - \xi/\xi_0} \tag{24}$$

where $\xi_0$ is the random close packing limit, close to 0.637 [7,8]; if we combine that asymptotic behaviour with our results we can recover a satisfactory enough behaviour over the whole density range.

Now it is well known that the Carnahan & Starling equation of state is obtained by rounding to the nearest integer the virial coefficients obtained by numerical techniques, i.e.

$$\frac{PV}{Nk_BT} \cong 1 + 4\xi + 10\xi^2 + 18\xi^3 + 28\xi^4 + 40\xi^5 + ... \qquad (25)$$

or

$$\frac{PV}{Nk_BT} \cong \frac{1 + \xi + \xi^2 - \xi^3}{(1-\xi)^3} \qquad (26)$$

The virial coefficients are thus simply given by $B_p = (p-1)(p+2)$.

If in equ.(25) we round our coefficients to the nearest decade and assume that the series is continuing beyond $p = 6$ we obtain

$$\frac{PV}{Nk_BT} \cong 1 + 4\xi + 10\xi^2 + 20\xi^3 + 30\xi^4 + 40\xi^5 + ... \qquad (27)$$

or

$$\frac{PV}{Nk_BT} \cong 1 + 4\xi + \frac{10\xi^2}{(1-\xi)^2} = \frac{1 + 2\xi + 3\xi^2 + 4\xi^3}{(1-\xi)^2} \qquad (28)$$

The virial coefficients for $p \geq 3$ appear as the successive multiples of $B_3 = 10$. That very simple equation of state suffers from the same drawback as Carnahan & Starling's: it exhibits a pole only for $\xi = 1$ whereas a pole is expected at the random close packing limit. Now, although it may look rather crude, it is nearly as good from a practical point of view.

**Conclusion**

We have proposed an original approximation of the correlation terms which allows a simple calculation of the virial coefficients of a fluid, which is explicitly carried through in the hard sphere case. The virial coefficients calculated within our approximation scheme are in rather good agreement with the known computed values up to order 6, i.e. at low density; if we extrapolate those results for higher orders we obtain a simple equation of state in good agreement with the experimental data for the stable fluid phase.